\documentstyle[12pt]{article}
\begin{document}
\def\t{\theta}
\def\ov{\overline}
\def\a{\alpha}
\def\b{\beta}
\def\g{\gamma}
\def\wt{\widetilde}

\addtolength{\unitlength}{-0.5\unitlength}

\newsavebox{\wfrontsq}\savebox{\wfrontsq}(75,75){\begin{picture}(75,75)
\thicklines\multiput(0,0)(0,75){2}{\line(1,0){75}}
\multiput(0,0)(75,0){2}{\line(0,1){75}}\end{picture}}

\newsavebox{\wleftsq}\savebox{\wleftsq}(37.5,112.5)
{\begin{picture}(37.5,112.5)
\thicklines\multiput(0,37.5)(0,75){2}{\line(1,-1){37.5}}
\multiput(0,37.5)(37.5,-37.5){2}{\line(0,1){75}}\end{picture}}

\newsavebox{\wupsq}\savebox{\wupsq}(112.5,37.5)
{\begin{picture}(112.5,37.5)
\thicklines\multiput(0,37.5)(75,0){2}{\line(1,-1){37.5}}
\multiput(0,37.5)(37.5,-37.5){2}{\line(1,0){75}}\end{picture}}

\newsavebox{\bfrontsq}\savebox{\bfrontsq}(75,75){\begin{picture}(75,75)
\thicklines\multiput(0,0)(0,75){2}{\line(1,0){75}}
\multiput(0,0)(75,0){2}{\line(0,1){75}}\put(0,0){\line(1,1){75}}
\multiput(0,50)(50,-50){2}{\line(1,1){25}}
\end{picture}}

\newsavebox{\bleftsq}\savebox{\bleftsq}(37.5,112.5)
{\begin{picture}(37.5,112.5)
\thicklines\multiput(0,37.5)(0,75){2}{\line(1,-1){37.5}}
\multiput(0,37.5)(37.5,-37.5){2}{\line(0,1){75}}
\put(37.5,0){\line(-1,3){37.5}}
\multiput(12.5,25)(25,25){2}{\line(-1,3){12.5}}
\end{picture}}

\newsavebox{\bupsq}\savebox{\bupsq}(112.5,37.5)
{\begin{picture}(112.5,37.5)
\thicklines\multiput(0,37.5)(75,0){2}{\line(1,-1){37.5}}
\multiput(0,37.5)(37.5,-37.5){2}{\line(1,0){75}}
\put(0,37.5){\line(3,-1){112.5}}
\multiput(62.5,0)(25,25){2}{\line(-3,1){37.5}}
\end{picture}}

\begin{flushright} May, 1993
\end{flushright}
\vspace{3cm}
\begin{center}
{\bf Elliptic solution for modified tetrahedron equations}\\
\vspace{1cm}
V.V. Mangazeev, Yu.G. Stroganov\\
\vspace{1cm}
{\it Institute for High Energy Physics,}\\
{\it Protvino, Moscow Region, Russia}
\end{center}
\vspace{1cm}
\begin{flushleft}
{\bf Abstract}
\end{flushleft}
As is known, tetrahedron equations lead to the commuting family
of transfer-matrices and provide the integrability
of corresponding three-dimensional lattice models. We present
the modified version of these equations which give the commuting
family of more complicated two-layer transfer-matrices. In the static
limit we have succeeded in constructing the solution of these
equations in terms of elliptic functions.

\newpage

\section{Introduction}

Unlike the theory of solvable two-dimensional lattice models of
statistical mechanics the theory of three-dimensional models is
quite poor. Up to recently two-state Zamolodchikov model
\cite{Z1}, \cite{Z2}, \cite{B} has been the only nontrivial example of three-
dimensional solvable model. In recent paper \cite{BB1} Bazhanov and Baxter
have generalized Zamolodchikov model for an arbitrary number of states. The
Boltzmann weights of their model satisfy the three-dimensional
star-star equation, tetrahedron equations and possess
some remarkable symmetries under the rotations of the elementary
cube of the lattice \cite{KMS1}, \cite{BB2}, \cite{KMS2}.

In his first paper \cite{Z1} Zamolodchikov have constructed his solution
of the tetrahedron equations in the so called ''static'' limit. He interpreted
his solution in terms of scattering amplitudes of straight strings.
The static limit corresponds to the scattering of strings with zero
velocities. Further the static solution was generalized for arbitrary
velocities of straight strings and all string amplitudes were parameterized by
the trigonometric functions depending on tetrahedron angles \cite{Z2}.

However, the first unpublished version of paper \cite{Z1} have contained
the model in the static limit, whose weight functions were parameterized in
terms of elliptic functions. It turned out, that these weight functions
satisfy only some part of tetrahedron equations. The numerical computer tests,
carried out by Bazhanov in 1980, showed that one can satisfy each
concrete tetrahedron equation by changing the signs of weight functions,
but all efforts to choose the signs of weight functions so that
all tetrahedron equations were satisfied were futile.

In this paper we show that the sign factors of weight functions of elliptic
Zamolodchikov model can be chosen in a such way that these weight
functions satisfy the modified system of tetrahedron equations.
On obeying these equations we have succeeded in constructing the
three-dimensional integrable model with commuting family of two-layer
transfer matrices following the ideology of paper \cite{KS}.

\section{Weight functions of the model}

The first variant of Zamolodchikov model of straight strings corresponds
to the statistical three-dimensional model on  a cubic lattice with spin
variables lying on the faces of the lattice. Baxter (see, for example
\cite{B}) has shown that one can reformulate Zamolodchikov model as an
interaction-round-cube model with spins in the sites of the lattice.
Hereafter we will follow the ideology and notations of paper \cite{B}.

The weight function $W(a|efg|bcd|h;\t_1,\t_2,\t_3)$ of the model depends on
the values of eight spin variables, surrounding one elementary cube of the
lattice (see Fig. 1), and three spectral parameters
$\t_1$, $\t_2$, $\t_3$, satisfying the static limit condition:
\begin{equation}
                       \t_1+\t_2+\t_3=\pi.                       \label{1}
\end{equation}
\begin{picture}(600,265)
\put(0,50){
\begin{picture}(500,200)
\multiput(140,0)(120,0){2}{\line(0,1){120}}
\multiput(140,0)(0,120){2}{\line(1,0){120}}
\multiput(140,0)(0,120){2}{\line(-1,1){60}}
\put(80,180){\line(1,0){120}}\put(80,180){\line(0,-1){120}}
\put(200,180){\line(1,-1){60}}
\multiput(200,180)(0,-20){6}{\line(0,-1){12}}
\multiput(80,60)(20,0){6}{\line(1,0){12}}
\multiput(255,5)(-30,30){2}{\line(-1,1){20}}
\multiput(140,0)(120,0){2}{\circle*{15}}
\multiput(140,120)(120,0){2}{\circle*{15}}
\multiput(80,60)(120,0){2}{\circle*{15}}
\multiput(80,180)(120,0){2}{\circle*{15}}
\put(300,80){\large $=\quad W(a|e,f,g|b,c,d|h;\theta_1,\theta_2,\theta_3)$}
\put(150,10){$e$}\put(270,10){$d$}\put(150,100){$a$}
\put(270,100){$f$}\put(212,186){$b$}\put(92,190){$g$}
\put(92,65){$c$}\put(212,65){$h$}
\end{picture}
}
\put(320,0){\Large\bf Fig. 1}
\end{picture}
\vspace{0.5cm}

\noindent
Hereafter all spins take the values $0,\>1$
and we will imply that all calculations over the spin variables
are carried out {\it modulo} 2.

The $W$ function satisfies the following ''recoloring'' symmetry relations:
\begin{eqnarray}
W(a|e,f,g|b,c,d|h)&=&W(\ov a|e,f,g|\ov b,\ov c,\ov d|h)=\nonumber\\
=W(a|\ov e,\ov f,\ov g|b,c,d|\ov h)&=&
W(\ov a|\ov e,\ov f,\ov g|\ov b,\ov c,\ov d|\ov h),             \label{2}
\end{eqnarray}
where we omit the dependence on spectral parameters, $\ov a=1-a$,
$\ov b=1-b$, etc.
Taking into account symmetry relations (\ref{2}) we have $64$ different
weight functions.

Define the following combinations of spin variables $a,e,f,g,b,c,d,h$:
\begin{eqnarray}
j_1=a+b+e+h,&& j_2=a+c+f+h,\quad j_3=a+d+g+h,\nonumber\\
m_1=e+h,&&m_2=f+h,\quad\quad\quad\quad m_3=g+h,                   \label{3}
\end{eqnarray}
and their combinations:
\begin{equation}
l_1=j_2+j_3,\quad l_2=j_1+j_3,\quad l_3=j_1+j_2.                   \label{4}
\end{equation}
All spin variables in formulas (\ref{3}-\ref{4}) take the values $0,1$ and
the weight function $W$ in the static limit is defined by formula:
\begin{equation}
W(a|e,f,g|b,c,d|h;\t_1,\t_2,\t_3)=(-1)^{\Phi(j_1,j_2,j_3,m_1,m_2,m_3)}
{s_1^{j_1}s_2^{j_2}s_3^{j_3}\over c_1^{l_1}c_2^{l_2}c_3^{l_3}}
k^{j_1j_2j_3},                                                      \label{5}
\end{equation}
where we use the following notations:
\begin{equation}
s_i=\biggl[sn({\t_i K\over\pi})\biggr]^{1/2},\quad
c_i=\Biggl[{cn({\t_i K\over\pi})\over dn({\t_i K\over\pi})}\Biggr]^{1/2},\quad
i=1,2,3,                                                            \label{6}
\end{equation}
$sn$, $cn$, $dn$ are elliptic functions of modulus $k$, $K$ is the complete
elliptic integral of the first kind (see, for example \cite{GR}).

If we take spectral parameters $\t_i$, satisfying  the condition
$0\le\t_i\le\pi$, $i=1,2,3$, then all values of the elliptic functions in
(\ref{6}) are non-negative, and we choose the positive signs of all square
roots in formulas (\ref{6}).

The sign factor $\Phi$ in formula (\ref{5}) is defined by the following
expression:
\begin{eqnarray}
&&\Phi(j_1,j_2,j_3,m_1,m_2,m_3)=
j_1m_2m_3+j_2m_3m_1+j_3m_1m_2+\nonumber\\
&&+n_1(j_1m_2m_3+j_2m_3m_1+j_3m_1m_2+
j_1j_2m_3+j_2j_3m_1+j_3j_1m_2)+\nonumber\\
&&+n_2j_1j_2j_3+n_3(j_1+j_2+j_3)+n_4(m_1+m_2+m_3),                \label{7}
\end{eqnarray}
where $n_1$, $n_2$, $n_3$, $n_4$ are arbitrary integer numbers.

Hereafter we will suppose static limit condition (\ref{1}) to be valid
and use the following notation:
\begin{equation}
W(a|e,f,g|b,c,d|h;\t_1,\t_2)\equiv W(a|e,f,g|b,c,d|h;\t_1,\pi-\t_1-\t_2,\t_2).
                                                                  \label{8}
\end{equation}

\section{Modified tetrahedron equations}

Let us consider the following equation
\begin{eqnarray}
\sum_d    W(a_1|c_1c_2c_3|b_1b_2b_3|d;\t_1,\t_2)
      \ov W(c_1|a_2b_3b_2|dc_6c_4|b_4;\t_1,\t_2+\t_3)\times&&\nonumber\\
   \times W(b_3|c_4c_2d|b_1b_4a_3|c_5;\t_1+\t_2,\t_3)
      \ov W(d|b_4b_1b_2|c_3c_6c_5|a_4;\t_2,\t_3)=&&\nonumber\\
=\sum_d\ov W(d|a_2a_3a_4|c_5c_6c_4|b_4;\t_1,\t_2)
          W(a_1|dc_2c_3|b_1a_4a_3|c_5;\t_1,\t_2+\t_3)\times&&\nonumber\\
\times\ov W(c_1|a_2a_1b_2|c_3c_6d|a_4;\t_1+\t_2,\t_3)
          W(b_3|c_4c_2c_1|a_1a_2a_3|d;\t_2,\t_3),&&          \label{9}
\end{eqnarray}

\noindent
In appendix we will prove that the weight function $W$ given by formulas
(\ref{5}-\ref{8}) satisfies modified tetrahedron equations (\ref{9})
with
\begin{equation}
\ov W(a|e,f,g|b,c,d|h;\t_1,\t_2)=W(h|b,c,d|e,f,g|a;\t_1,\t_2).   \label{10}
\end{equation}

Note that additional weight functions $\ov W$ can be obtained from
weight function $W$ with the help of the so called $(\tau\rho)^3$
transformation (see \cite{KMS1},\cite{BB2}). For the Bazhanov-Baxter model,
the weight functions are invariant under $(\tau\rho)^3$ transformation,
$\ov W=W$,  and modified tetrahedron equations (\ref{9}) are reduced to
the usual tetrahedron equations in the static limit.

In our case only absolute values of weight functions have this invariance,
but weights $\ov W$ differ from $W$ by some sign factors. More explicitly,
for any fixed set of integers $n_1,\>n_2,\>,n_3\>,n_4$
(see formula (\ref{7}))
we can choose another set $n'_1,\>n'_2,\>n'_3,\>n'_4$ in a such way
that the weight function $\ov W$ is replaced by $W$ and vice versa.

It means that the weight functions $W$ and $\ov W$ satisfy also
another equation:
\begin{eqnarray}
\sum_d\ov W(a_1|c_1c_2c_3|b_1b_2b_3|d;\t_1,\t_2)
          W(c_1|a_2b_3b_2|dc_6c_4|b_4;\t_1,\t_2+\t_3)\times&&\nonumber\\
\times\ov W(b_3|c_4c_2d|b_1b_4a_3|c_5;\t_1+\t_2,\t_3)
          W(d|b_4b_1b_2|c_3c_6c_5|a_4;\t_2,\t_3)=&&\nonumber\\
=\sum_d   W(d|a_2a_3a_4|c_5c_6c_4|b_4;\t_1,\t_2)
      \ov W(a_1|dc_2c_3|b_1a_4a_3|c_5;\t_1,\t_2+\t_3)\times&&\nonumber\\
\times    W(c_1|a_2a_1b_2|c_3c_6d|a_4;\t_1+\t_2,\t_3)
      \ov W(b_3|c_4c_2c_1|a_1a_2a_3|d;\t_2,\t_3).&&          \label{11}
\end{eqnarray}

\section{Solvable model on three-dimensional lattice}

As is known, tetrahedron equations lead to the commuting family
of transfer-matrices and provide the integrability
of corresponding three-dimensional lattice models \cite{BS}, \cite{JM}.
In paper \cite{KS} it was shown that one can use the modified Yang-Baxter
equation for constructing the commuting family of two-layer transfer-matrices.
Similarly, we have succeeded in constructing a commuting family of two-layer
transfer matrices for the special model on three-dimensional cubic
lattice, where weight functions $W$ alternate with weight functions $\ov W$
in a checkerboard order in all directions. Note that the proof of this
fact demands that both equations (\ref{9}), (\ref{11}) should be valid.

Horizontal transfer-matrices $T(\t_1,\t_2)$ constructed with the help
of alternating weights $W(\t_1,\t_2)$ and $\ov W(\t_1,\t_2)$ (see Fig. 2),
satisfy the commutation relation:
\begin{equation}
[T(\t_1,\t_2),T(\t_1,\t_3)]=0.                                 \label{12}
\end{equation}
\begin{picture}(700,410)
\put(100,70){
\begin{picture}(600,300)
\multiput(150,0)(150,0){3}{\usebox{\wfrontsq}}
\multiput(150,75)(150,0){3}{\usebox{\bfrontsq}}
\multiput(225,0)(150,0){3}{\usebox{\bfrontsq}}
\multiput(225,75)(150,0){3}{\usebox{\wfrontsq}}
\multiput(112.5,0)(-75,75){2}{\usebox{\wleftsq}}
\multiput(112.5,75)(-75,75){2}{\usebox{\bleftsq}}
\multiput(75,37.5)(-75,75){2}{\usebox{\bleftsq}}
\multiput(75,112.5)(-75,75){2}{\usebox{\wleftsq}}
\multiput(112.5,150)(150,0){3}{\usebox{\bupsq}}
\multiput(187.5,150)(150,0){3}{\usebox{\wupsq}}
\multiput(75,187.5)(150,0){3}{\usebox{\wupsq}}
\multiput(150,187.5)(150,0){3}{\usebox{\bupsq}}
\multiput(37.5,225)(150,0){3}{\usebox{\bupsq}}
\multiput(112.5,225)(150,0){3}{\usebox{\wupsq}}
\multiput(0,262.5)(150,0){3}{\usebox{\wupsq}}
\multiput(75,262.5)(150,0){3}{\usebox{\bupsq}}
\end{picture}}
\put(310,0){\Large\bf Fig. 2}
\end{picture}
\vspace{0.5cm}

\noindent
More explicitly, we have shown that the weight factor corresponding
to the composite cube constructed from four ''elementary'' weights $W$
and four ''elementary'' weights $\ov W$ (see Fig. 3) satisfies the
tetrahedron equations of mixed type with spin variables placed in
the corner sites, in the middles of edges and in the centers of all faces
of the composite cube.

\begin{picture}(600,400)
\put(150,70){
\begin{picture}(300,300)
\multiput(0,100)(0,100){3}{\line(1,-1){100}}
\multiput(100,0)(-50,50){3}{\line(0,1){200}}
\multiput(100,0)(0,100){3}{\line(1,0){200}}
\multiput(200,0)(100,0){2}{\line(0,1){200}}
\multiput(300,200)(-100,0){2}{\line(-1,1){100}}
\multiput(0,300)(50,-50){2}{\line(1,0){200}}
\multiput(100,0)(100,0){3}{\circle*{15}}
\multiput(100,100)(100,0){3}{\circle*{15}}
\multiput(100,200)(100,0){3}{\circle*{15}}
\multiput(50,250)(100,0){3}{\circle*{15}}
\multiput(0,300)(100,0){3}{\circle*{15}}
\multiput(0,200)(50,-50){2}{\circle*{15}}
\multiput(0,100)(50,-50){2}{\circle*{15}}
\multiput(50,50)(50,50){2}{\line(-1,3){50}}
\multiput(25,75)(50,50){2}{\line(-1,3){25}}
\multiput(50,100)(50,50){2}{\line(-1,3){25}}
\multiput(100,100)(100,-100){2}{\line(1,1){100}}
\multiput(150,100)(100,-100){2}{\line(1,1){50}}
\multiput(100,150)(100,-100){2}{\line(1,1){50}}
\multiput(200,200)(50,50){2}{\line(-3,1){150}}
\multiput(150,200)(50,50){2}{\line(-3,1){75}}
\multiput(175,225)(50,50){2}{\line(-3,1){75}}
\multiput(289,50)(0,1){3}{\line(1,0){100}}
\multiput(289,150)(0,1){3}{\line(1,0){100}}
\multiput(289,50)(0,1){3}{\line(2,1){20}}
\multiput(289,50)(0,1){3}{\line(2,-1){20}}
\multiput(289,150)(0,1){3}{\line(2,1){20}}
\multiput(289,150)(0,1){3}{\line(2,-1){20}}
\put(400,140){\Large $W$}\put(400,40){\Large $\overline W$}

\end{picture}
}
\put(320,0){\Large\bf Fig. 3}
\end{picture}

\noindent
The proof of this fact is based on the repeated
16-fold application of modified tetrahedron equations (\ref{9}), (\ref{11})
and will be published in a more detailed version of our paper.

\section{Conclusion}

In this paper we have formulated the statistical model on a three-dimensional
cubic lattice with Boltzmann weights parameterized in terms of elliptic
functions. Further it is naturally to ask about the existence of the deviation
from the static limit. It seems that the broken $(\tau\rho)^3$-invariance
is a principal feature of this model. The point is the following:
the limit $k\to0$ of our model looks very similar to the static limit
of the Zamolodchikov model but differs by some sign factors. Hence,
the generalization of the static elliptic solution (if it exists)
cannot be obtained by the deformation of the ''full'' Zamolodchikov solution.

Also it will be interesting to calculate the statistical sum of this model
and to investigate its critical behaviour.  And at last it is naturally to ask
about possible a generalization of this model for the number of states
$N>2$ by analogy with the Bazhanov-Baxter model.

\section{Appendix}

In this appendix we give the sketch of the proof of modified tetrahedron
equations (\ref{9}) for the weight functions $W$ and $\ov W$
given by formulas (\ref{5}-\ref{8}) and (\ref{10}).
 First let us introduce some useful notations.
Instead of spins $a_i$, $b_i$, $c_i\in Z_2$ we will use new spin variables
with values $\pm1$ and denote them
by Greek letters:
\begin{eqnarray}
&&\a_i=1-2a_i,\quad \b_i=1-2b_i,  \quad i=1,\ldots,4,\nonumber\\
&&\g_i=1-2c_i,\quad i=1,\ldots,6.                                 \label{13}
\end{eqnarray}
Define also nine combinations of spins (\ref{13}):
\begin{eqnarray}
\mu_1=\a_1\a_2\b_1\b_4,&&\mu_2=\a_1\a_4\b_3\b_4,\quad
\mu_3=\a_3\a_4\b_2\b_3,\nonumber\\
\rho_1=\a_1\b_1\g_1,\quad\rho_2=\a_1\b_3\g_3,&&
\rho_3=\a_3\b_3\g_5,\quad\rho_4=\a_2\b_2\g_4,\nonumber\\
\nu_1=\a_1\b_2\g_2,&&\nu_3=\a_4\b_1\g_6.                      \label{14}
\end{eqnarray}

Also let us introduce the following notations:
\begin{equation}
\wt s_i=k^{1/2}sn({\t_i K\over\pi}),\quad
\wt c_i=k^{1/2}{ cn({\t_i K\over\pi})\over dn({\t_i K\over\pi})},
\quad i=1,2,3,                                                     \label{15}
\end{equation}
\begin{eqnarray}
\wt s_{12}=k^{1/2}sn({(\t_1+\t_2) K\over\pi}),&&
\wt s_{23}=k^{1/2}sn({(\t_2+\t_3) K\over\pi}),\nonumber\\
\wt c_{12}=k^{1/2}
{cn({(\t_1+\t_2) K\over\pi})\over dn({(\t_1+\t_2) K\over\pi})},&&
\wt c_{23}=k^{1/2}
{cn({(\t_2+\t_3) K\over\pi})\over dn({(\t_2+\t_3) K\over\pi})},\nonumber\\
\wt s_{123}=k^{1/2}sn({(\t_1+\t_2+\t_3) K\over\pi}),&&
\wt c_{123}=k^{1/2}{cn({(\t_1+\t_2+\t_3) K\over\pi})
\over dn({(\t_1+\t_2+\t_3)K\over\pi})}.                           \label{16}
\end{eqnarray}
Substituting formulas (\ref{5}-\ref{8}), (\ref{10}) for $W$ and $\ov W$
functions in modified tetrahedron equations (\ref{9}) and taking into account
formulas (\ref{13}-\ref{16}) we obtain
\newpage

\begin{eqnarray}
k^{[\nu_1\rho_1\rho_2{(\mu_3-1)\over2}{(1+\mu_2\nu_1\nu_3\rho_2\rho_4)\over4}+
\nu_1\rho_3\rho_4{(1-\mu_1)\over2}{(1-\mu_2\mu_3\nu_1\nu_3\rho_2\rho_4)\over4}
]}\times\phantom{1111111}&&\nonumber\\
(-1)^{n_2{(1-\mu_1)\over2}{(1-\mu_2)\over2}{(1-\mu_3)\over2}}
\bigl(\wt c_1^{{(1-\mu_3)\over2}\nu_1\rho_2}
\wt c_3^{{(\mu_1-1)\over2}\nu_1\rho_4}
\bigr)^{(1+\mu_1\mu_2\nu_1\nu_3\rho_2\rho_4)\over2}\times\phantom{1111}
&&\nonumber\\
\bigl(\wt c_2^{{(1-\mu_1\mu_2\mu_3)\over2}\nu_1\rho_1}
\wt c_{123}^{{(1-\mu_2)\over2}\nu_3\rho_1}
\wt c_{12}^{{(\mu_2\mu_3-1)\over2}\nu_1}
\wt c_{23}^{{(\mu_1\mu_2-1)\over2}\nu_3}\bigr)^{
(1+\mu_3\nu_1\nu_3\rho_1\rho_3)\over2}\times\phantom{1111}&&\nonumber\\
\bigl(\wt s_{12}^{{(1+\mu_1\mu_2)\over2}\rho_1\rho_2}
\wt s_{23}^{{(1+\mu_2\mu_3)\over2}\rho_2\rho_3}
\wt s_2^{-{(1+\mu_2)\over2}\rho_2}
\wt s_{123}^{{(1+\mu_1\mu_2\mu_3)\over2}\rho_4}\bigr)^{
(1-\mu_1\mu_2\mu_3\rho_1\rho_2\rho_3\rho_4)\over2}\times\phantom{111}
&&\nonumber\\
\bigl[1+\mu_2\wt s_1^{{(1+\mu_1)\over2}\rho_1}
\wt s_2^{{(1+\mu_2)\over2}\rho_2}
\wt s_3^{{(1+\mu_3)\over2}\rho_3}
\wt s_{123}^{-{(1+\mu_1\mu_2\mu_3)\over2}\rho_4}
\wt c_{12}^{{(1-\mu_2\mu_3)\over2}\nu_1}
\wt c_{23}^{{(1-\mu_1\mu_2)\over2}\nu_3}\times\phantom{11}&&\nonumber\\
(-1)^{(1+n_2)(1-\mu_3\nu_1\nu_3\rho_1\rho_3)
(1-\mu_1\mu_2\nu_1\nu_3\rho_2\rho_4)\over4}
k^{\rho_1[\nu_1\rho_2
{(1+\mu_1\rho_1\rho_2\rho_3\rho_4)\over4}+\mu_2
\nu_3\rho_4{(1+\mu_3\rho_1\rho_2\rho_3\rho_4)\over4}]}\bigr]&&\nonumber\\
=1+\mu_2\wt s_1^{{(1+\mu_1)\over2}\rho_1}
\wt s_2^{{(1+\mu_2)\over2}\mu_1\mu_3\rho_1\rho_3\rho_4}
\wt s_3^{{(1+\mu_3)\over2}\rho_3}
\wt s_{123}^{-{(1+\mu_1\mu_2\mu_3)\over2}\rho_1\rho_2\rho_3}\times
\phantom{11111}&&\nonumber\\
\wt c_{12}^{{(1-\mu_2\mu_3)\over2}\mu_2\rho_1\rho_3\nu_3}
\wt c_{23}^{{(\mu_1\mu_2-1)\over2}\mu_3\nu_1\rho_1\rho_3}
(-1)^{(1+n_2)(1-\mu_3\nu_1\nu_3\rho_1\rho_3)
(1-\mu_1\mu_2\nu_1\nu_3\rho_2\rho_4)\over4}\times\phantom{11}&&\nonumber\\
k^{\rho_4[\nu_1\rho_3{(1+\mu_3\rho_1\rho_2\rho_3\rho_4)\over2}+
\mu_2\mu_3\nu_3\rho_1{(1+\mu_1\rho_1\rho_2\rho_3\rho_4)\over2}]}.
\phantom{111111111111111}&&                                       \label{17}
\end{eqnarray}

It appears that these relations depend on only nine independent
spin variables (\ref{14}). So we have $2^9=512$ separate relations
corresponding to the different choices of spin variables.
We split these relations into the eight groups corresponding to the eight
possible choices of spins $\mu_i=\pm1$, $i=1,2,3$.
Note that some of these equations have the form
\begin{equation}
A+B=A+B,\quad A-A=B-B,                                             \label{18}
\end{equation}
where $A$ and $B$ are the products of elliptic functions. We will call
these equations as trivial ones.

 First let us consider the case
\begin{equation}
\mu_1=1,\quad\mu_2=1,\quad\mu_3=1.                                \label{19}
\end{equation}
It is easily to show that for such choice of $\mu_i$
we have only eight nontrivial relations which can be written in the
following form:
\begin{equation}
\wt s_{12}^{\rho_1\rho_2}\wt s_{23}^{\rho_2\rho_3}(1+
\wt s_1^{\rho_1}\wt s_2^{\rho_2}
\wt s_3^{\rho_3}\wt s_{123}^{\rho_1\rho_2\rho_3})=
\wt s_1^{\rho_1}\wt s_3^{\rho_3}+
\wt s_2^{\rho_2}\wt s_{123}^{\rho_1\rho_2\rho_3}.                \label{20}
\end{equation}

Detailed analysis shows that for all other choices of spins $\mu_i$,
$i=1,2,3$ equations (\ref{17}) are reduced to equation (\ref{20}).
Note that in some cases we should specify the arguments of elliptic
functions in equation (\ref{20}) to some fixed values (for example,
$\t_2+\t_3=\pi$) and use the quasiperiodic conditions for elliptic
functions $sn(x)$, $cn(x)$ and $dn(x)$.

It still remains to check eight relations (\ref{20}) for arbitrary
spins $\rho_1$, $\rho_2$ and $\rho_3$ taking the values $\pm1$.
Note that any choice of spins $\rho_i$ in (\ref{20}) can be reduced to
the case $\rho_1=1$, $\rho_2=1$ and $\rho_3=1$ with the help
of the following formula:
\begin{equation}
sn(iK'+x)={1\over k\>sn(x)},                                     \label{21}
\end{equation}
where $K'$ is the complete integral of the first kind of the
complementary modulus $k'=(1-k^2)^{1\over2}$.

Hence, we should prove equation (\ref{20})  for the only choice
$\rho_1=1$, $\rho_2=1$ and $\rho_3=1$.
This equation can be easily checked by using of addition theorems
for elliptic functions. It completes our proof of modified
tetrahedron equations (\ref{9}).


\begin{thebibliography}{**}

\bibitem{Z1}
A.B. Zamolodchikov, Zh. Eksp. Teor. Fiz. {\bf79} (1980) 641-664
[English trans.: JETP {\bf 52} (1980) 325-336].

\bibitem{Z2}
A.B. Zamolodchikov, Commun. Math. Phys. {\bf79} (1981) 489-505.

\bibitem{B}
R.J. Baxter, Commun. Math. Phys. {\bf88} (1983) 185-205.

\bibitem{BB1}
V.V. Bazhanov, R.J. Baxter, J. Stat. Phys. {\bf 69} (1992) 453-485.

\bibitem{KMS1}
R.M. Kashaev, V.V. Mangazeev, Yu.G. Stroganov, Int. J. Mod. Phys. {\bf A8}
(1993) 587-601.

\bibitem{BB2}
V.V. Bazhanov, R.J. Baxter, ``Star-Triangle Relation for a Three-Dimensional
Model'', J. Stat. Phys., 1993 (in press).

\bibitem{KMS2}
R.M. Kashaev, V.V. Mangazeev, Yu.G. Stroganov, Int. J. Mod. Phys. {\bf A8}
(1993) 1399-1409.

\bibitem{KS}
R.M. Kashaev, Yu.G. Stroganov, ``Generalized Yang-Baxter Equation'',
to be published.

\bibitem{GR}
I.S. Gradshtein, I.M. Ryzhik, ``Tables of Integrals, Series and Products,
Academic Press, New York, 1965.

\bibitem{BS}
V.V. Bazhanov, Yu.G. Stroganov, Teor. Mat. Fiz. {\bf52} (1982) 105-113
[English trans.: Theor. Math. Phys. {\bf52} (1982) 685-691].

\bibitem{JM}
M.T. Jaekel, J.M. Maillard, J. Phys. {\bf A15} (1982) 1309.

\end{thebibliography}
\end{document}